\documentclass[a4paper]{jpconf}
\usepackage{graphicx}
\usepackage{amssymb}

\usepackage{citesort}

\begin{document}

\title{Di-photon Higgs decay in the MSSM with explicit CP violation}

\author{S Hesselbach$^{1,}$\footnote[99]{Speaker}, S Moretti$^{1,2}$,
  S Munir$^1$ and P Poulose$^{1,3}$}

\address{$^1$ School of Physics \& Astronomy, University of Southampton,
  Highfield, Southampton SO17~1BJ, UK}
\address{$^2$ Laboratoire de Physique Th\'eorique, Universit\'e
  Paris--Sud, F--91405 Orsay Cedex, France}
\address{$^3$ Physics Department, IIT Guwahati, Assam, INDIA - 781039}

\ead{s.hesselbach@phys.soton.ac.uk}

\begin{abstract}
The Minimal Supersymmetric Standard Model (MSSM) with explicit CP
violation is studied with the help of the di-photon decay channel of the
lightest neutral Higgs boson.
Effects of CP violation, entering via the scalar/pseudo-scalar mixing
at higher order as well as through the Higgs-sfermion-sfermion
couplings at tree-level, are analyzed in the MSSM with and without
light sparticles.
A light stop may have a strong impact on the decay width and Branching
Ratio (BR) of the decay process $H_1\rightarrow \gamma\gamma$, whereas
other light sparticles have only little influence.
In some regions of the MSSM parameter space with large CP-violating
phase $\phi_\mu\sim 90^\circ$ 
a light stop can change the BR by more than 50\%.
\end{abstract}

\section{Introduction}

Supersymmetry (SUSY) is one of the most favoured scenarios 
of new physics and will be searched for in all possible ways at the
upcoming Large Hadron Collider (LHC) at CERN.
In the Minimal Supersymmetric Standard Model (MSSM) many parameters
can well be complex and thus explicitly break CP invariance inducing
CP violation also in the Higgs sector beyond Born approximation
\cite{Pilaftsis:1998dd}.
After elimination of unphysical phases and imposing universality conditions at
the unification scale two independent phases remain,
the phase $\phi_\mu$ of the higgsino mass term $\mu$ and
a common phase $\phi_{A_f}$ of the soft trilinear Yukawa couplings $A_f$
in the sfermion sector \cite{Pilaftsis:1999qt}.
Despite the fact that the SUSY phases may be severely constrained by
bounds on the Electric Dipole Moments (EDMs), these constraints are rather
model dependent and may be evaded in scenarios with heavy first and second
generation sfermions, due to cancellations among various contributions
to the EDMs or due to additional contributions from lepton flavour
violating terms in the Lagrangian, for a review see e.g.~\cite{Olive:2005ru}.

In this contribution
we study the di-photon decay mode, $H_1\rightarrow \gamma\gamma$,
of the lightest neutral Higgs boson $H_1$, which involves
direct, i.e\ leading, effects of the SUSY phases through couplings of
the $H_1$ to SUSY particles in the loops
as well as indirect, i.e.\ sub-leading, effects through the
scalar/pseudo-scalar mixing yielding the Higgs mass-eigenstate $H_1$.
In scenarios with heavy SUSY particles, where the CP violation enters
solely through the scalar/pseudo-scalar mixing, the SUSY CP phases can
result in a strong suppression of the branching ratio (BR) of the
decay $H_1\rightarrow \gamma\gamma$ as well as of the rate of the
combined production and decay
process $gg \to H_1 \to \gamma\gamma$ \cite{Choi:2001iu}.
Here, we summarize the results of \cite{Moretti:2007th,Hesselbach:2007en} 
focusing especially on the effects of light SUSY particles
on the decay $H_1\rightarrow \gamma\gamma$.
The analysis of the full production and decay process at the LHC is
postponed to a forthcoming publication \cite{fullprocess}.

\section{Di-photon Higgs decay in CP-violating MSSM}

In order to analyze the Higgs decays in the CP-violating MSSM we have
used the publicly available \textsc{Fortran} code \textsc{CPSuperH}
\cite{Lee:2003nta}, version 2, which calculates the mass spectrum and
decay widths of all Higgs bosons along with their couplings to SM and
SUSY particles.
The leading terms in the CP-violating scalar/pseudo-scalar mixing in
the Higgs sector are proportional to $\mathrm{Im}(\mu A_f)$, hence we
assume $\phi_{A_f} = 0$ and analyze the effects of nonzero $\phi_\mu$
in the following.

A random parameter space scan to study the general
behaviour of the $\mathrm{BR}(H_1 \to \gamma\gamma)$ for non-zero
$\phi_\mu$ has revealed that about 50\% deviations are possible for
$M_{H_1}$ around 104 GeV for $\phi_\mu=100^\circ$, and an average of 30\%
deviation occurs over the mass range 90--130 GeV.
Furthermore, a strong impact
of a light stop $\tilde{t}_1$ on the BR has been established
\cite{Moretti:2007th}.
In \cite{Hesselbach:2007en} these results have been consolidated by
analyzing the details at the matrix element level including the phase
dependence of the Higgs mixing matrix elements and of the respective
couplings and by performing a more thorough study of the dependence on
the SUSY parameters.

\begin{figure}[t]
\includegraphics[width=18pc]{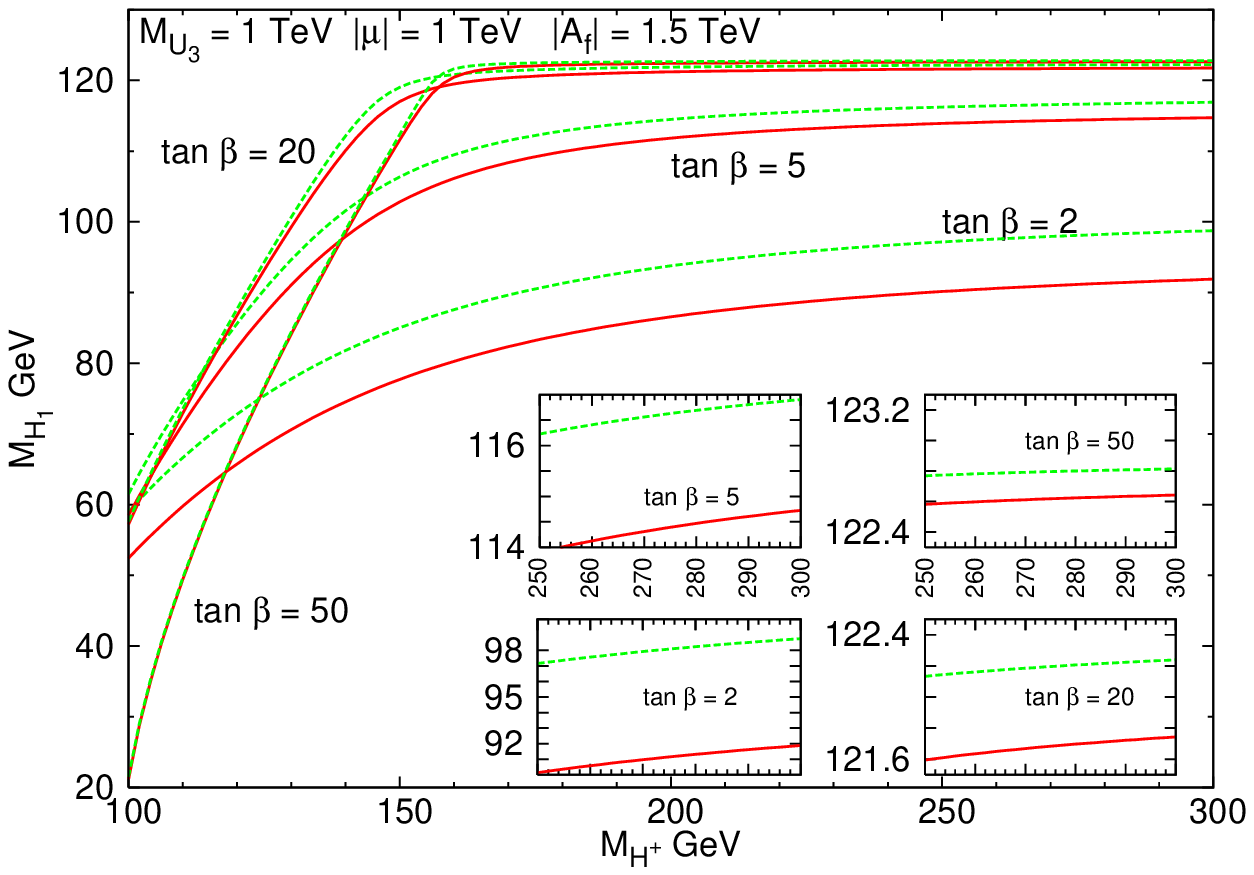}
\hspace{\fill}
\includegraphics[width=18pc]{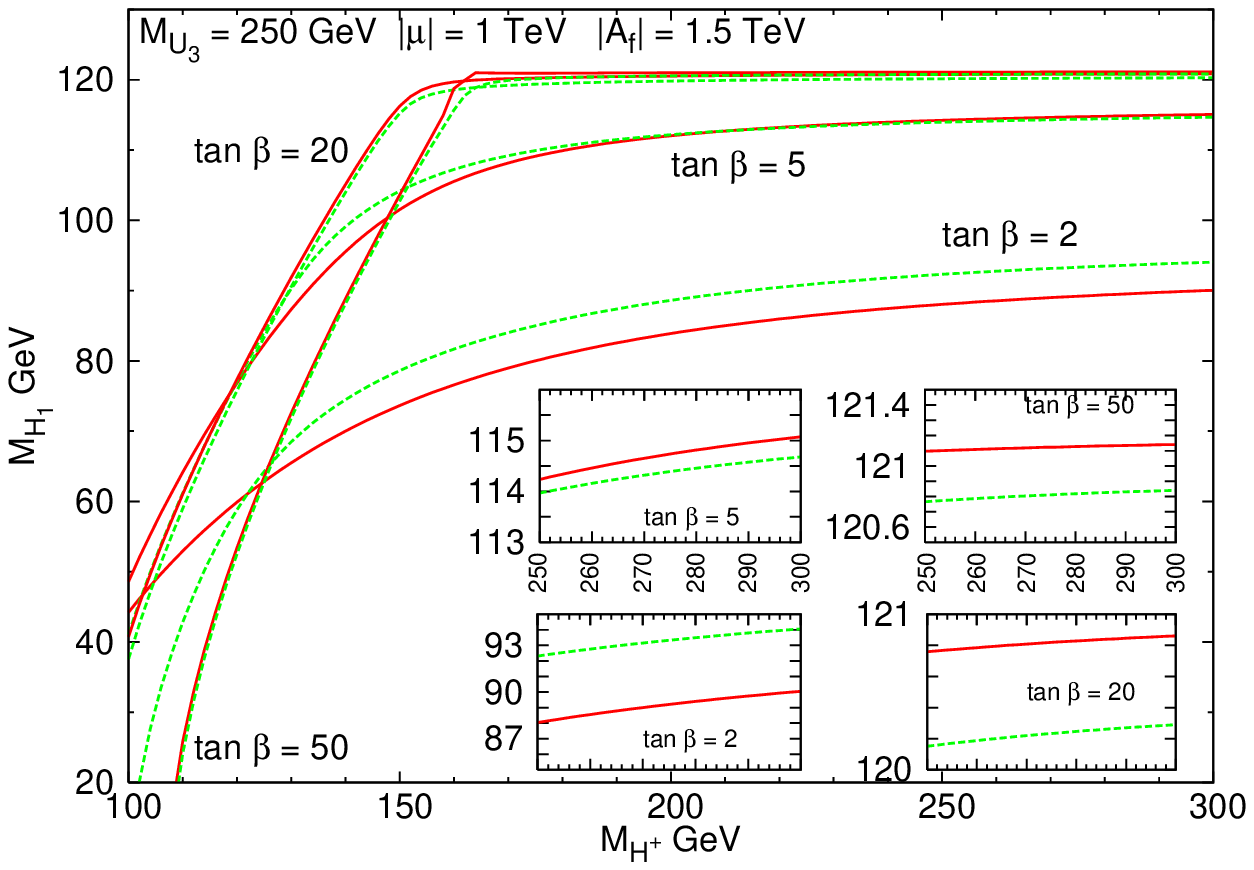}
\caption{\label{fig:Higgsmass}Mass of the lightest neutral Higgs boson
$H_1$ for $\phi_\mu=0^\circ$ (solid, red line)
and $\phi_\mu=90^\circ$ (dashed, green line)
with $|A_f|=1.5$ TeV, $|\mu|=1$~TeV and different values of $\tan\beta$.
In the left plot all sparticles are heavy ($\sim 1$ TeV) for
$M_{\tilde U_3}=M_{\tilde Q_3}=M_{\textrm{SUSY}} = 1$~TeV, whereas
in the right plot a light stop ($\sim 200$ GeV) is present for
$M_{\tilde U_3}=250$ GeV and $M_{\tilde Q_3}=M_{\rm SUSY}=1$~TeV.}
\end{figure}

In figure~\ref{fig:Higgsmass} the mass of $H_1$ is shown as a function
of the mass of the charged Higgs boson ($M_{H^+}$)
for $\phi_\mu=0^\circ, 90^\circ$, 
$\tan\beta=2,5,20,50$ and the two cases
$M_{\tilde U_3}=M_{\tilde Q_3}=M_{\textrm{SUSY}} = 1$~TeV, where all
SUSY particles are heavy, and $M_{\tilde U_3}=250$ GeV and
$M_{\tilde Q_3}=M_{\rm SUSY}=1$~TeV,
where a light $\tilde{t}_1$ with mass $\sim 200$ GeV is present.
While in the low $\tan\beta$ case the mass shift induced by the change
in $\phi_ \mu$ from $0^\circ$ to $90^\circ$
is about 10\%, in the case of $\tan\beta=20$ or above it is about
1\% or less.
The sudden shift in the dependence of $M_{H_1}$ on $M_{H^+}$
around $M_{H^+}=150$ GeV is due to a cross over in the Higgs mass
eigenstates at that point, where all three neutral Higgs states are
approximately degenerate in mass around $120$~GeV.

\begin{figure}[t]
\includegraphics[width=18pc]{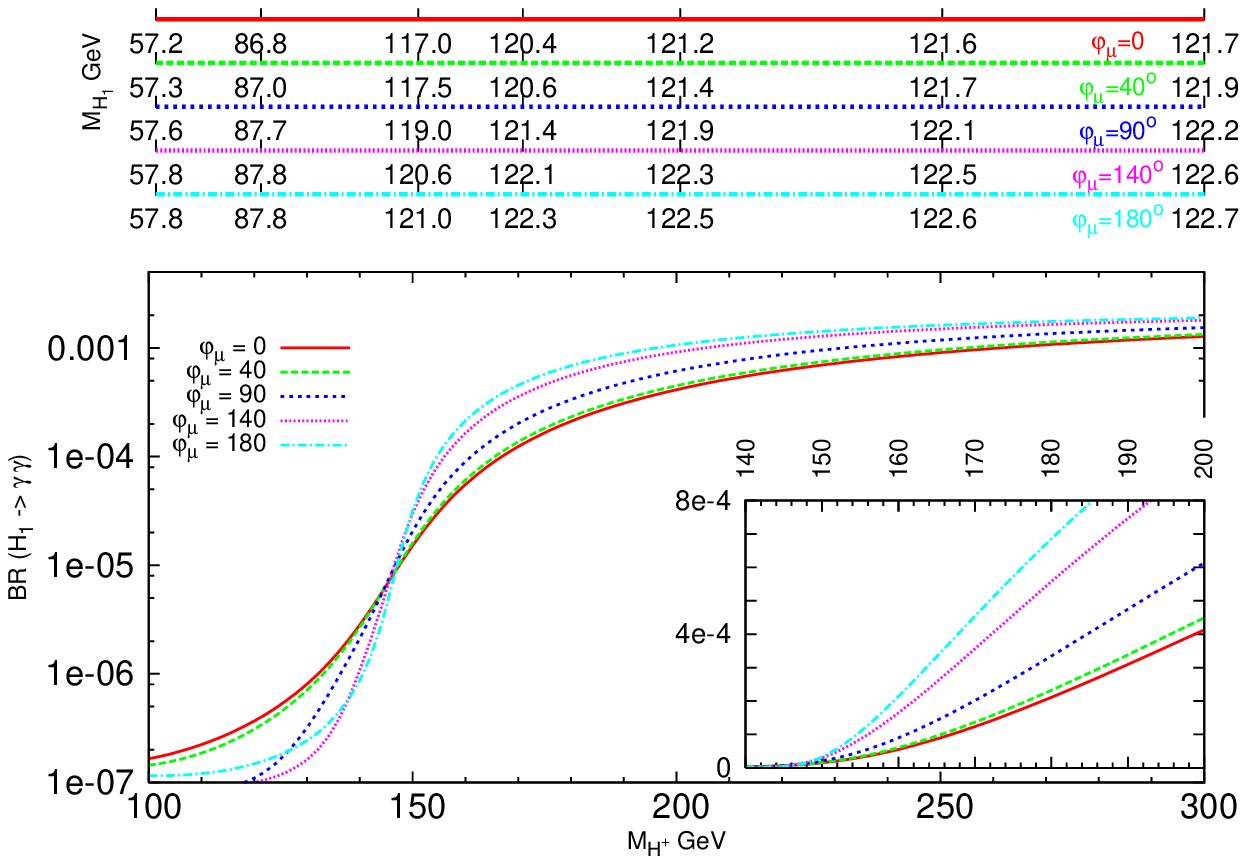}
\hspace{\fill}
\includegraphics[width=18pc]{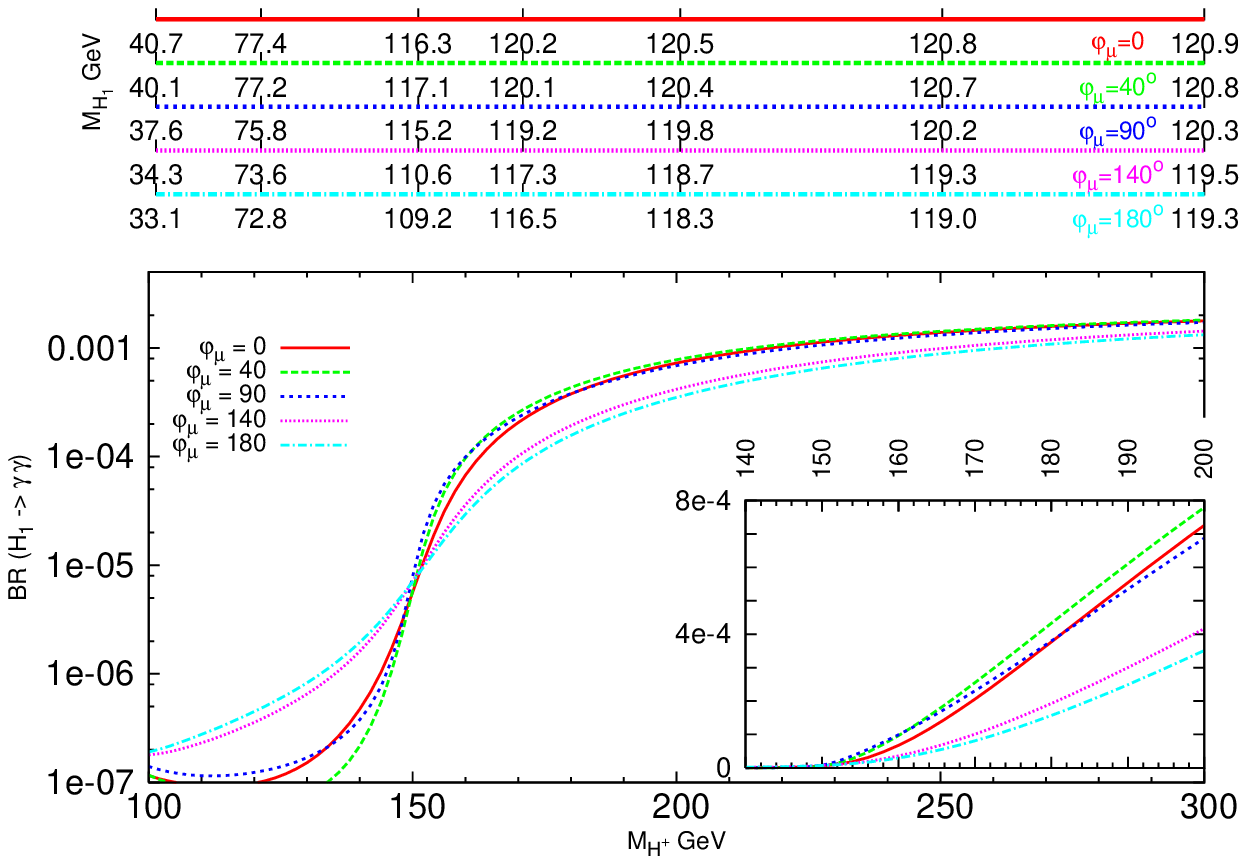}
\caption{\label{fig:BR}
BR of $H_1\rightarrow \gamma\gamma$
for $|A_f|=1.5$ TeV, $|\mu|=1$ TeV and $\tan\beta=20$.
Values of $M_{H_1}$ corresponding to representative points on the
$M_{H^+}$ axis
are indicated on the horizontal lines above separately for the values of
$\phi_\mu$ used.
The left plot corresponds to the case with $M_{\tilde U_3}=1$~TeV
(no light SUSY particles),
while the right plot corresponds to the case with
$M_{\tilde U_3}=250$~GeV (a light stop is present).}
\end{figure}

Figure~\ref{fig:BR} shows the BR($H_1\rightarrow \gamma\gamma$)
for five representative $\phi_\mu$ values between $0^\circ$ and $180^\circ$
as a function of $M_{H^+}$ for the two cases
$M_{\tilde U_3}=1$~TeV (all SUSY particles heavy) and
$M_{\tilde U_3}=250$ GeV (light $\tilde{t}_1$).
The respective values of $M_{H_1}$ are indicated separately on the horizontal
lines above for each $\phi_\mu$ value.
Again the cross over point in the Higgs mass eigenstates at
$M_{H^+} \sim 150$ GeV is clearly visible.
Below this point the BRs are very small and there is a strong $\phi_\mu$
dependence of $M_{H_1}$, hence our analysis is not relevant in this
parameter region.
Above $M_{H^+} \sim 150$ GeV with $M_{H_1} \gtrsim 115$~GeV
the $\phi_\mu$ dependence of $M_{H_1}$ is within the expected
experimental uncertainty
and the BR is large enough to be important
for the LHC Higgs search.
In scenarios with heavy SUSY particles (left plot) the BR increases with
increasing $\phi_\mu$ leading to a 50\% increase for
$\phi_\mu=90^\circ$ at $M_{H^+}\sim 200$ GeV.
This $\phi_\mu$ dependence is caused mainly by the $\phi_\mu$ dependence 
of the $H_1$ couplings to $W^\pm$ bosons and $t$ and $b$ quarks which appear
in the loop-induced decay $H_1\rightarrow \gamma\gamma$.
When a light $\tilde{t}_1$ is present the additional $\phi_\mu$
dependence in the stop sector causes a considerable change of the
$\phi_\mu$ dependence of the BR.
First the BR increases again with increasing $\phi_\mu$ up to a maximum 
for some value of $\phi_\mu$ around $40^\circ$, beyond which, however, the
BR decreases to about 50\% at $\phi_\mu = 180^\circ$.

Concerning the dependence on other SUSY parameters we have found
that a smaller value of $|A_f|$ considerably changes the $\phi_\mu$
dependence of the BR in scenarios with light $\tilde{t}_1$, whereas
a smaller $|\mu|$ value leads generally to a smaller $\phi_\mu$
dependence. Other light SUSY particles have only a negligible effect
on the BR.

\section{Summary}

We have analyzed the BR of the di-photon decay of the
lightest Higgs boson in the CP-violating MSSM with complex $\mu$
parameter.
We have found that the strong $\phi_\mu$ dependence of the BR
considerably changes in the presence of a light scalar top
$\tilde{t}_1$.
In general, the BR may be increased or decreased for a non-zero
$\phi_\mu$ depending on the SUSY parameter point.

\section*{References}
\bibliography{hgamgamcpv}

\end{document}